\documentclass[prd,reprint,groupedaddress,nofootinbib]{revtex4-2}

\usepackage{graphicx}
\usepackage{dcolumn}
\usepackage{bm}
\usepackage{mathtools}
\usepackage{amssymb,amsmath}
\usepackage[protrusion=true,expansion=true]{microtype}
\usepackage[dvipsnames]{xcolor}
\usepackage{microtype}
\usepackage{siunitx}
\usepackage{multirow}
\usepackage[colorlinks=true,linkcolor=blue,citecolor=blue,urlcolor=blue]{hyperref}

\usepackage{ragged2e}
\newcolumntype{L}[1]{>{\raggedright\arraybackslash}p{#1}}
\newcolumntype{C}[1]{>{\centering\arraybackslash}p{#1}}
\newcolumntype{R}[1]{>{\raggedleft\arraybackslash}p{#1}}
\newcolumntype{J}[1]{>{\justifying\arraybackslash}p{#1}}

\definecolor{webred}{rgb}{0.75,0,0}

\definecolor{chapter_green}{HTML}{006400}

\newcommand{\unbound}[1]{\textcolor{gray}{#1}}

\begin{document}

\title{Structure of open-flavour four-quark states in the charm and bottom region}

\author{Joshua Hoffer$^{1,2}$}
\email[]{joshua.hoffer@theo.physik.uni-giessen.de}
\author{Gernot Eichmann$^3$}
\email[]{gernot.eichmann@uni-graz.at}
\author{Christian S. Fischer$^{1,2}$}
\email[]{christian.fischer@theo.physik.uni-giessen.de}

\affiliation{$^1$Institut für Theoretische Physik, Justus-Liebig-Universität Gießen, 35392 Gießen, Germany}
\affiliation{$^2$Helmholtz Forschungsakademie Hessen für FAIR (HFHF), GSI Helmholtzzentrum
für Schwerionenforschung, Campus Gießen, 35392 Gießen, Germany}
\affiliation{$^3$Institute of Physics, University of Graz, NAWI Graz, Universitätsplatz 5, 8010 Graz, Austria}

\date{\today}

\begin{abstract}
We present quantitative results for masses and the internal structure of four-quark states with two heavy quarks, 
i.e. $QQ'\bar{q}\bar{q}'$ with $Q,Q'\in \{c,b\}$ and $q,q' \in \{u,d,s\}$, and $J^{P} \in {1^+}$. 
The composition of these states in terms of meson-meson and diquark-antidiquark pairs, extracted from a relativistic 
four-body Faddeev-Yakubowski equation, is dynamically determined from underlying QCD forces. 
We find states at energy levels in very good agreement with lattice QCD and, where available, with experimental states. 
Their internal structure, most notably between the $T^+_{cc}$, $T_{bc}$ and $T^-_{bb}$, show significant and sizeable variations. 
\end{abstract}


\maketitle

\section{Introduction}\label{intro} 
It took almost 40 years from the introduction of the notion of `exotic' four-quark mesons \cite{GellMann1964} until the 
first such candidate, the $\chi_{c1}(3872)$, was identified by the BELLE collaboration \cite{Choi2003}. Since then, many more 
exotic meson candidates were discovered by Belle (II), \textit{BABAR}, BES III and the LHCb experiments in the strange, charm and bottom 
energy region. While some four-quark states have similar quantum numbers and even masses as conventional mesons, many have unique
signatures such as carrying an electric charge or conventionally forbidden decay patterns, see 
Refs.~\cite{Chen2016,Esposito2017,Lebed2017,Ali2017,Guo2018,Olsen2018,Liu2019,Brambilla2020} for recent review articles. 

At the time of writing, all observed hidden-flavour four-quark states with quark content $Q,Q'\in \{c,b\}$ and 
$q,q' \in \{u,d,s\}$  are resonances that decay predominantly via the strong interaction. 
The decay widths vary widely from a few to a few hundred \unit{\MeV} \cite{ParticleDataGroup:2024}. By contrast, 
open-flavour four-quark states with two heavy quarks and two light antiquarks\footnote{In this work we consider
the strong interaction only. Consequently $QQ'\bar{q}\bar{q}'$ and $\bar{Q}\bar{Q}'qq'$ are equivalent.}, i.e. $QQ'\bar{q}\bar{q}'$, are expected to form stable bound states {for isospin $I=0$}, provided the mass of 
the heavy quark pair is 
sufficiently large and the mass of the light antiquarks sufficiently small
\cite{Ballot1983,Zouzou1986,J.Lipkin1986,Heller1987,Manohar1993}. The heavy $QQ'$ pair then behaves almost like a 
point-like, colour-antitriplet antiquark and the resulting binding mechanism is similar as in heavy-light baryons. 
Indeed, theoretical calculations in recent years find a deeply bound $bb\bar{u}\bar{d}$ four-quark state,
called $T_{bb}^-$, with a binding energy around $-$(100 -- 200) MeV with respect to the $BB^\ast$ threshold, see e.g.
\cite{Francis2017,Eichten2017,Braaten2021,Hudspith2023,Alexandrou2024} and Refs. therein. Its charm-quark 
counterpart $T_{cc}^+$ with $cc\bar{u}\bar{d}$ is the only open-flavour state that has been identified experimentally
\cite{Aaij2022,Aaij2022a}. With a binding energy of $-273(61)$ \unit{\keV} with respect to the $D^0 D^{\ast +}$ threshold, 
this state is extremely shallow and has a narrow decay width of only 410(165) \unit{\keV}.

The internal structure of open-flavour four-quark states is highly debated in the literature.
Provided two-body forces are dominant, there are two distinct possibilities: the four (anti)quarks may form heavy-light meson-meson pairs or may arrange in (heavy-heavy)(light-light) diquark-antidiquark pairs. These correlations in flavour
and colour space may then result in a spatial structure such as meson molecules \cite{Guo2018} or compact heavy diquarks
\cite{Eichten2017,Braaten2021}. For the latter case it has been argued that the formation of a corresponding light scalar 
diquark-antidiquark pair is an important binding mechanism for the four-quark state. 

In order to settle these questions, it is crucial to develop approaches without 
\textit{a priori} assumptions on the internal 
structure of these states. This is the main topic of this work. We report on results from functional methods, i.e. 
from the four-body Faddeev-Yakubowski equation and underlying Dyson-Schwinger equations. In a series of works
\cite{Heupel2012,Eichmann2016a,Santowsky2022} we have shown that the framework is capable of describing the resonant 
nature of the $f_0(500)$, together with its partners in the lightest scalar meson nonet, and qualitative properties 
of hidden-flavour heavy-light four-quark states \cite{Wallbott2019,Wallbott2020,Santowsky2022a,Hoffer2024}. Two-body 
clusters in the four-body states are dynamically generated by the underlying QCD interactions without prejudice or
assumptions on dominant meson-meson or diquark-antidiquark structures. As we will see in the course of this work, 
this is particularly important as the interplay of flavour structure and resulting symmetries generates different
dominant structures for different states. A further crucial, and novel, element  compared to previous works in
the functional approach is the inclusion of both  {\em attractive and repulsive} meson-meson and diquark-antidiquark
configurations. As it turns out, this is mandatory to render our calculation quantitative for   states with 
both $I(J^{P}) = 0(1^+)$ and $I(J^{P}) = 1(1^+)$. We furthermore develop the techniques to quantify the contributions 
of these configurations to the total norm of these states, thus offering important insights into their internal 
structure.

We briefly summarise our framework in the next section and discuss results for the spectrum and internal structure in 
Sec.~\ref{results}, before we conclude in Sec.~\ref{conclusions}. 
%
Further technical details are given in the Appendix.
%
%
%

\begin{figure}[t]
	\includegraphics[scale=0.38]{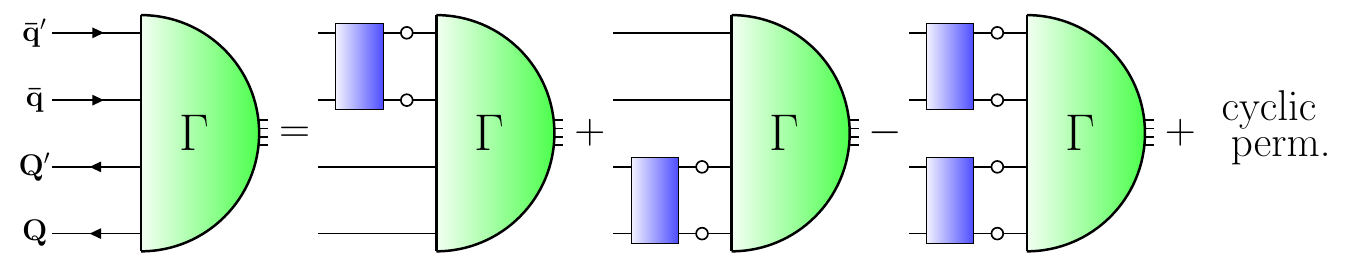}
	\caption{\label{fig: bse}Four-quark BSE for an open-flavour $QQ'\bar{q}\bar{q}'$ system in the $(12)(34)$ topology; the permutations $(13)(24)$ and $(14)(23)$ are not shown. The green half-circles represent the Bethe-Salpeter amplitudes, blue boxes the two-body interaction kernels and the circles fully dressed quark propagators.}
\end{figure}

\section{Four-quark Equation}\label{bse} 
The four-body Faddeev-Yakubowski equation is an exact homogeneous Bethe-Salpeter equation (BSE) in QCD.
Its solution provides the location of the mass pole in the complex energy plane, i.e., a real mass on the 
first Riemann sheet for a bound state below threshold or a complex (real) mass on higher sheets 
for resonances (virtual states). In compact notation it has the generic form 
\begin{align}\label{eq: four-quark bse}
\Gamma^{(4)} = K^{(4)} G_0^{(4)}\, \Gamma^{(4)}\, ,
\end{align}
where each multiplication corresponds to an integration over all loop momenta. The interaction kernel  
$K^{(4)}$  contains all possible two-, three- and four-body interactions, $\Gamma^{(4)}$ is the four-quark Bethe-Salpeter 
amplitude (BSA), and $G_0^{(4)}$ is the product of four dressed (anti)quark propagators; 
see \cite{Heupel2012,Eichmann2015,Wallbott2019,Hoffer2024} for details. Eq.~(\ref{eq: four-quark bse}) 
is  an eigenvalue equation, where the pole locations are identified by the eigenvalue $\lambda_i(P^2)$ 
of $K^{(4)}G_0^{(4)}$ satisfying $\lambda_i(P^2) = 1$. Here $i=0$ stands for the ground state and $i=1,2,\ldots$ for radially excited states.

The main focus of this work is to study the properties of open-flavour four-quark states in terms of their composition
in internal two-body clusters. 
As in our previous works, we therefore assume dominance of two-body-correlations and neglect irreducible 
three- and four-body interactions \cite{Eichmann2015,Wallbott2019,Wallbott2020,Santowsky2022a}. The interaction kernel then 
reads
\begin{align}\label{eq: kernel}
K^{(4)}G_0^{(4)} = \sum_{aa'}\,\left(K_a+K_a'-K_{aa'}\right)\, ,
\end{align}
with $a$ and $a'$ denoting interactions between quark-(anti)quark pairs. The last term in Eq.~(\ref{eq: kernel}) is necessary 
to avoid overcounting \cite{Huang1975,Kvinikhidze1992,Heupel2012}. For our case of an open-flavour four-quark state 
$QQ'\bar{q}\bar{q}'$, with $Q,\,Q'\in\{b,c\}$ and $q,\,q'\in\{u,d,s,c\}$, the resulting equation is shown diagrammatically 
in Figure~\ref{fig: bse}. The pair $aa'$ can be one of three combinations $(12)(34)$, $(13)(24)$ and $(14)(23)$ corresponding 
to the possible two-body interaction topologies: One (heavy-heavy)(light-light) diquark-antidiquark $(QQ')(\bar{q}\bar{q}')$ 
and two heavy-light meson-meson topologies $(Q\bar{q})(Q'\bar{q}')$ and $(Q\bar{q}')(Q'\bar{q})$. 
{For the two-body interaction we employ the rainbow-ladder truncation, where the quark-(anti-)quark kernel
is absorbed into an effective gluon exchange.
This is the leading order in a systematic truncation scheme and has been successfully used to determine spectra of mesons, baryons and four-quark states 
in the past, see \cite{Eichmann2016,Eichmann2020} for reviews.
The explicit form of the interaction is discussed 
%
in App. \ref{app: two-body}.
%
%
%
}

\begin{table}
\setlength{\tabcolsep}{0.36em} 
\centering
\begin{tabular}{p{6mm}C{6mm}||cc|c|cc|c}
	$I(J^{P})$ &  & \multicolumn{6}{c}{Physical components}\\[4pt]
	&  & \multicolumn{2}{c}{$\mathbf{1}\otimes\mathbf{1}$} & $\bar{\mathbf{3}}\otimes\mathbf{3}$ &
	\multicolumn{2}{c}{$\mathbf{8}\otimes\mathbf{8}$} & $\mathbf{6}\otimes\bar{\mathbf{6}}$\\[4pt]
	&  & $f_0$ & $f_1$ & $f_2$ & $f_3$ & $f_4$ & $f_5$  \\[4pt]                                                                  
	\hline\hline&&&&&&&\\[-7pt] 
	{$0(1^+)$} 
	&$bb\bar{n}\bar{n}$ & $B B^\ast$ & $B^\ast B^\ast$ & $A_{bb}S$ & $B B^\ast$ & $B^\ast B^\ast$ & $S_{bb}A$\\[4pt]
	&$bc\bar{n}\bar{n}$ & $B D^\ast$ & $B^\ast D$      & $A_{bc}S$ & $B D^\ast$ & $B^\ast D$      & $S_{bc}A$\\[4pt]
	&$cc\bar{n}\bar{n}$ & $D D^\ast$ & $D^\ast D^\ast$ & $A_{cc}S$ & $D D^\ast$ & $D^\ast D^\ast$ & $S_{cc}A$\\[4pt]
	&$bb\bar{s}\bar{s}$ & $B_s B_s^\ast$ & $-$ & $A_{bb}A_{ss}$ & $B_s B_s^\ast$ & $-$ & $-$\\[4pt]
	&$bc\bar{s}\bar{s}$ & $B_s D_s^\ast$ & $B_s^\ast D_s$      & $S_{bc}A_{ss}$ & $B_s D_s^\ast$ & $B_s^\ast D_s^\ast$      & $A_{bc}S_{ss}$           \\[4pt]
	&$cc\bar{s}\bar{s}$ & $D_s D_s^\ast$ & $-$ & $A_{cc}A_{ss}$ & $D_s D_s^\ast$ & $-$ & $-$\\[4pt]
	\hline&&&&&&&\\[-7pt] 
	{$1(1^+)$} 
	&$bb\bar{q}\bar{q}$ & $B B^\ast$ & $-$             & $A_{bb}A$ & $B B^\ast$ & $-$             & $-$      \\[4pt]
	&$bc\bar{q}\bar{q}$ & $B D^\ast$ & $B^\ast D$      & $S_{bc}A$ & $B D^\ast$ & $B^\ast D^\ast$      & $A_{bc}S$\\[4pt]
	&$cc\bar{q}\bar{q}$ & $D D^\ast$ & $-$             & $A_{cc}A$ & $D D^\ast$ & $-$             & $-$      \\[4pt]
\end{tabular}
\caption{\label{tab: components} Physical content of the BS amplitude for quark and colour configurations investigated in this 
work, with $n \in \{u,d\}$. States with $\bar{c}\bar{c}$ are analogous to those with $\bar{s}\bar{s}$ and not shown explicitly.  
Scalar and axialvector diquarks are denoted by $S$ and $A$ respectively, with the subscript denoting the heavy-quark content 
of the diquark. We  grouped the physical components according to their attractive and repulsive colour structure;
$f_0,f_1,f_2,f_3, f_4,f_5$ correspond to the dressing functions for that particular component and colour channel. 
Empty slots in meson-meson channels are not populated for physics reasons, while those in diquark channels are forbidden by symmetry constraints.}
\end{table}

For given quantum numbers $J^{P}$, the four-quark Bethe-Salpeter amplitude can be written as
\begin{align}\label{eq: bsa_general}
\Gamma^{(\mu)}(P,p,k,q) = \Gamma_{\rm{D}}^{(\mu)}(P,p,k,q) \otimes \Gamma_{\rm{C}} \otimes \Gamma_{\rm{F}}\, ,
\end{align}
with the total momentum $P$ of the four-quark state and relative momenta $p,k,q$ between the four (anti)quarks. 
$\Gamma_{\rm{D}}$, $\Gamma_{\rm{C}}$ and $\Gamma_{\rm{F}}$ are the Dirac, colour and flavour parts, respectively. 
The Lorentz index $\mu$ applies to $J=1$ states. 

Following the arguments in~\cite{Hoffer2024}, 
we reduce the full BSA, initially containing $256$ Dirac structures for $J=0$ and $768$ for $J=1$,   
to the physical BSA expressed in terms of the six dominant internal two-body clusters for the different interaction topologies given in Table~\ref{tab: components}:
\begin{align}\label{eq: bsa_physical}
	\Gamma^{(\mu)}(S_0,a,s) \approx\sum_{j=0}^5f_j(S_0)\,\tau_j^{(\mu)}(S_0,a,s)\otimes \tau_j^\mathrm{C}\otimes \tau_j^\mathrm{F}\, .
\end{align}
The dressing functions $f_j$  depend on the permutation-group invariant momentum scale $S_0=(p^2+q^2+k^2)/4$.
The Dirac-Lorentz tensors $\tau_j^{(\mu)}$  already include the internal two-body pole structures for the physical components given in Table~\ref{tab: components} and therefore depend on $S_0$ but also on the doublet variables
$a=\sqrt{3}(q^2-p^2)/4$ and $s=(p^2+q^2-2k^2)/4$.
The $\tau_j^{\rm{F}}$ are the flavour tensors {for details on momentum dependencies see \cite{Eichmann2015,Wallbott2020}}.
The $\tau_j^{\rm{C}}$ are the colour singlet structures for each interaction topology,
where we include the attractive $\mathbf{1}\otimes\mathbf{1}$ and repulsive $\mathbf{8}\otimes\mathbf{8}$ color components for the meson-meson
topologies and the attractive $\bar{\mathbf{3}}\otimes\mathbf{3}$ and repulsive $\mathbf{6}\otimes\bar{\mathbf{6}}$ 
colour tensors for the diquark-antidiquark topology (for details see Supplemental Material in \cite{Wallbott2020}).
 
It is important to realize that Eq.~\eqref{eq: bsa_physical} does not make any assumptions on the internal spatial distribution of the four-quark
state. All four topologies, $\mathbf{1}\otimes\mathbf{1}$, $\mathbf{8}\otimes\mathbf{8}$, $\bar{\mathbf{3}}\otimes\mathbf{3}$ and $\mathbf{6}\otimes\bar{\mathbf{6}}$, decode structure in colour and flavour space only. It is thus a purely dynamical question 
related to the details of the interactions between the (anti)quarks which of these topologies dominate, and which internal structure 
and spatial distribution follows from this dominance. 

{The procedure we employ to extract the masses of the four-quark states is described in App.~B of Ref.~\cite{Hoffer2024}.
We calculate the eigenvalues $\lambda_i(P^2)$ of the four-quark BSE and determine the masses $M_i$ from intersections where 
the condition $\lambda_i(P^2 = -M_i^2) = 1$ is satisfied. If the states lie above thresholds, we analytically continue the 
eigenvalues above the thresholds. At present we cannot reliably extract the widths of the resonances, which would require 
knowledge of the eigenvalues for complex  $P^2$  on the first sheet above the threshold (using contour deformations) and 
a subsequent analytic continuation to higher Riemann sheets. This procedure has been successfully employed in two-body
systems~\cite{Eichmann2019,Santowsky2022} but its implementation in the four-body equation remains the subject of future work.}

Note that results for spectra of open-flavour four-quark states have already been published by some of us in Ref.~\cite{Wallbott2020}.
These have been obtained with attractive color forces only and employing an extrapolation method for the eigenvalue curves 
(second order polynomials) that at the time was considered reliable with errors given in Ref.~\cite{Wallbott2020}. In 
the course 
of this work, we updated their results using improved extrapolation methods (amongst those higher order polynomials). Corresponding 
results are presented in appendix \ref{app: att_only}. While we found that the extrapolation of many states is indeed insensitive
to this improvement of methods, the masses of some states do change, e.g., the mass of the 
$T_{cc}^+$. We have carefully checked
that our improved method is much more robust and are therefore confident to state that the results of this work for open-flavour 
states supersede those of Ref.~\cite{Wallbott2020}.

\section{Results}\label{results} 

\begin{figure}
\centering
\includegraphics[width=1\columnwidth]{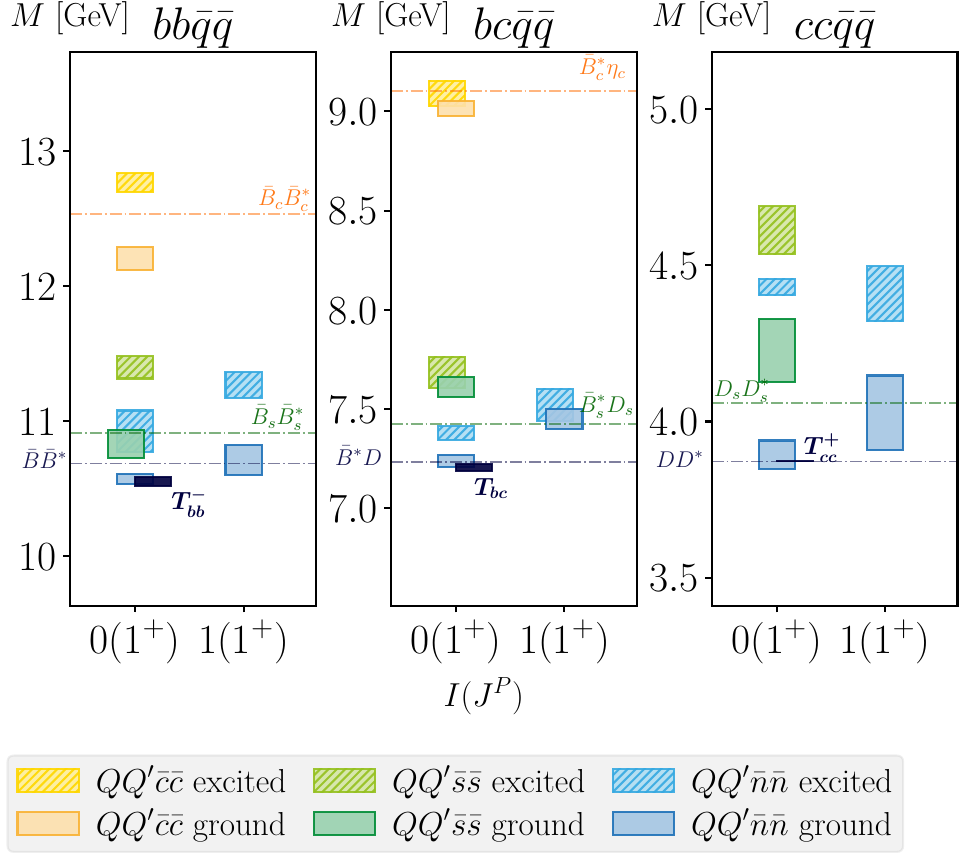}
\caption{\label{fig: equal_QQ} Spectrum for the open-flavour $J^P=1^+$ states. 
From left to right we show the spectra of open-bottom $bb\bar{q}\bar{q}$, open-bottom-charm $bc\bar{q}\bar{q}$ and open-charm $cc\bar{q}\bar{q}$. 
The heights of the boxes represent the error of the extracted masses. As a reference, we plotted the lowest relevant thresholds 
(see Table~\ref{tab: components}) for the respective four-quark systems (colour-coded according to the box colour). The black box 
for the  $T_{cc}^{+}$ is the only experimentally known state at the time of writing, whereas the boxes for the $T_{bb}^-$ and $T_{bc}$ are the 
averaged theoretical results.}
\end{figure}

\begin{figure}
	\centering	
	\includegraphics[width=1\columnwidth]{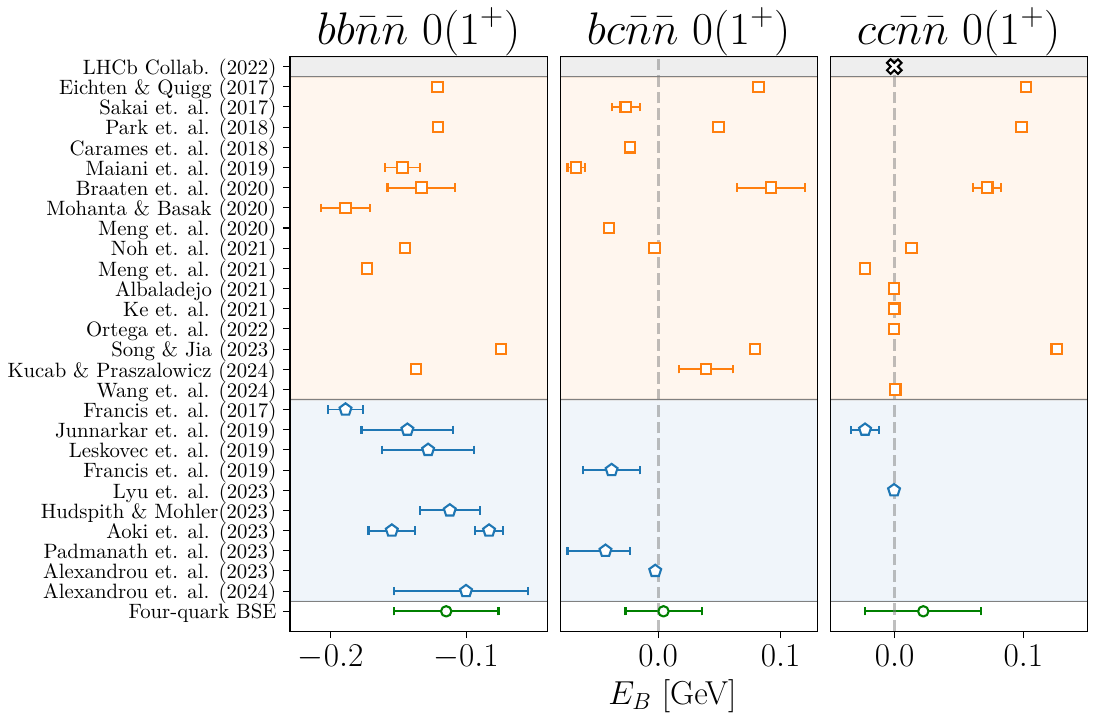} \includegraphics[width=1\columnwidth]{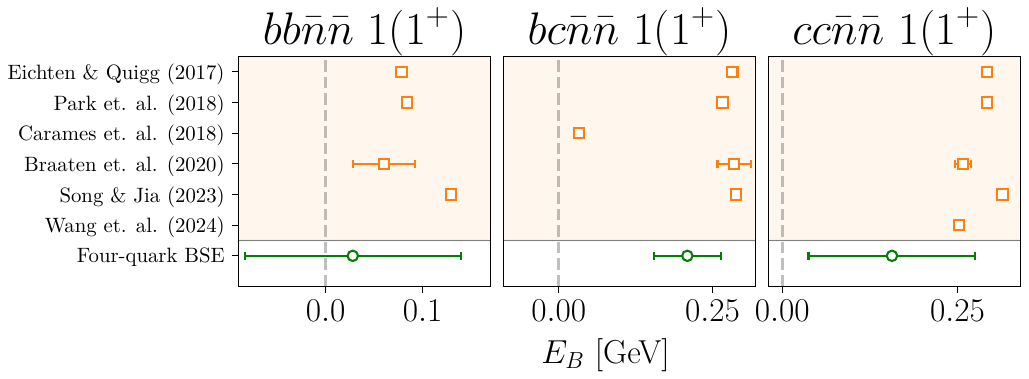} \\[-2mm]
	\caption{\label{fig: lit_comp_j1}Comparison of theoretical and (if available) experimental  results for the binding energy 
		$E_B$ of the $J^P=1^+$ ground states. The results in green are from this work. 
		The blue points are lattice results~\cite{Francis2017,Francis2019,Junnarkar2019,Leskovec2019,Lyu2023,Aoki2023,Hudspith2023,
		Padmanath2023,Alexandrou2024,Alexandrou2024a} 
		and the orange ones from phenomenology~\cite{Eichten2017,Sakai2017,Park2019,Carames2019,Maiani2019,Mohanta2020,Braaten2021,
			Meng2021,Noh2021,Meng2022,Albaladejo2022,Ke2022,Ortega2023,Song2023,Wang2024,Kucab2024}. 
			The grey dashed lines are the lowest meson-meson thresholds for the respective channel.}
		\vspace{-1mm}
\end{figure}

\subsection{Spectra and binding energies:}
Our results for the spectra of open-flavour ground and radially excited $QQ'\bar{q}\bar{q}'$ states  with  
$Q,Q' \in \{b,c\}$, $q=q' \in \{c,s,d,u\}$ and quantum numbers $I(J^P) = 0(1^+),\, 1(1^+)$ are shown 
in Fig.~\ref{fig: equal_QQ}. The corresponding masses are given 
%
in App.~\ref{app: masses}.  
%
We also plot the lowest relevant heavy-light meson-meson thresholds for the respective quantum numbers.
The respective meson masses are obtained from the two-body BSE and compiled in Table~\ref{tab: meson-masses} 
%
in App.~\ref{app: two-body}.
%

For the ground states in the $I(J^P)=0(1^+)$ channel, we find a clear hierarchy for the binding energies, i.e. the 
$bb\bar{n}\bar{n}$ (with $n \in \{u,d\}$) is deeply bound, the $bc\bar{n}\bar{n}$ is at best shallowly bound, and the $cc\bar{n}\bar{n}$ sits right at the $DD^\ast $ threshold.
This matches expectations from the literature: numerous studies including lattice QCD \cite{Bicudo:2012qt,Francis2017,Hudspith2023,Leskovec2019,Hudspith2020,Bicudo:2015vta,Bicudo:2015kna,Junnarkar2019,
	  Mohanta2020,Aoki2023,Alexandrou2024} identify a potentially deeply bound $T_{bb}^-$ as a $bb\bar{u}\bar{d}$ state and predict 
 less binding as the heavy quark mass decreases \cite{Alexandrou2024a,Hudspith2020,Meinel2022,Francis2019,Radhakrishnan2024,Padmanath:2022cvl,Cheung:2017tnt,Junnarkar2019,Lyu2023}. 
  This averaged value for the $T_{bb}^-$ from the literature
  is displayed in the left plot of Fig.~\ref{fig: equal_QQ} in black as a reference. For its binding energy,
  we find very good agreement  with the literature (top left in Fig.~\ref{fig: lit_comp_j1}).
  The same is true for the $T_{bc}$ (top center in Fig.~\ref{fig: lit_comp_j1}): Our result is slightly above the threshold but still more in accordance with recent lattice results \cite{Padmanath2023,Alexandrou2024a}. 
  For the experimentally known 
  $T_{cc}^+$ we find a binding energy of about $20\pm 50$ MeV with respect to the $DD^\ast $ threshold, which is in good agreement with 
  the shallow experimental state with binding energy $\delta m_{\mathrm{BW}} = -273(61)$ keV$/c^2$ \cite{Aaij2022} (top right in Fig.~\ref{fig: lit_comp_j1}).
  
  The interplay between attractive and repulsive forces is crucial for this agreement. Including only attractive 
  colour components yields (much) too strong binding. Depending on quantum numbers, adding the repulsive colour 
  components leads to upwards shifts of the masses in the range of $100-400$ MeV, see 
  App.~\ref{app: att_only} for details. Compared to other states, the 
  upwards shift of the $T_{cc}^+$ mass is quite modest with about $100$ MeV. 
  
  Regarding the $I(J^P)=1(1^+)$ states: It is well known e.g. from lattice QCD 
  \cite{Bicudo:2015vta,Bicudo:2015kna} that the repulsive forces are stronger for $I=1$ than for $I=0$ leading 
  to unbound states. Indeed, we also find these states to be unbound (bottom row in Fig.~\ref{fig: lit_comp_j1}). 
  
  For ground states with $\bar{s}\bar{s}$ quark content we see the same mass hierarchy from $QQ=bb$, $QQ=bc$ to 
  $QQ=cc$ as for the states with light quark content. In the $QQ=bb$ channel, the double strange ground state 
  is still below threshold, with a difference between double-light and double-strange ground states
  of about 250 MeV (compared to about 340 MeV for the $QQ=cc$ states).
  
  For the radial excitations, however, we find an interesting different pattern. We have chosen the  mass 
  ranges in the  spectra of Fig.~\ref{fig: equal_QQ} such that relative excitation energies with respect to
  the lowest threshold can be directly compared between the three channels. For the radial excitations, the 
  relative excitation energy is clearly smallest in the $QQ=bc$ channel. We  currently have no obvious explanation
  of this behaviour but suspect that the different symmetry structure of unequal heavy quarks $bc$ vs. equal heavy 
  quarks $cc$, $bb$ plays an important role. 
  
  Finally, we note that our heaviest states ($bb\bar{c}\bar{c}$, $bc\bar{c}\bar{c}$) are strongly bound. Whether
  this binding is too strong remains to be investigated. Lattice studies in the Born-Oppenheimer approximation 
  seem to exclude bound $bb\bar{c}\bar{c}$ states \cite{Bicudo:2015vta} but one may argue that the 
  Born-Oppenheimer approximation in this case is perhaps not entirely free of doubt.

\subsection{Internal structure:} 
\begin{figure}
	\centering
	\includegraphics[width=0.49\textwidth]{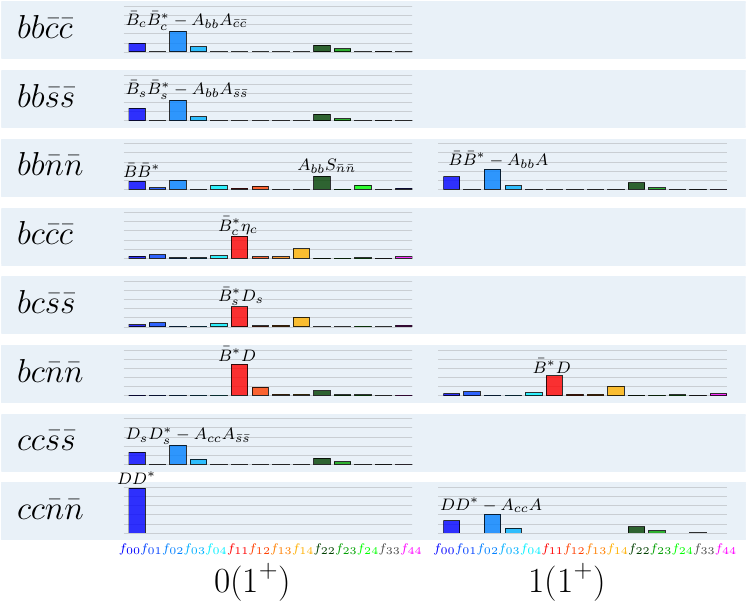}
	\caption{\label{fig: norm_contributions1} Norm contributions for the ground states in the axialvector $J^P = 1^+$ channel with $I=0,1$. 
		The results for the excited states (not shown) are very similar. 
		The distance between two horizontal lines in the bar plots is 20\%.
		Note that $I=1$ states with $\bar{s}\bar{s}$ and $\bar{c}\bar{c}$ do not exist.
	}
\end{figure}

The main novel element of our study compared to other approaches is the possibility to identify internal structures
in terms of clustering of meson and diquark pairs, namely through  norm contributions~\cite{Hoffer2024}:  
One takes the physical amplitude in Eq.~\eqref{eq: bsa_physical}, in this case consisting of six elements, 
and contracts it with its charge conjugate $\bar{\Gamma}^{(\mu)}$ (connected by quark propagators). 
This yields $6\times 6$ diagrams, 
where the diagonal elements are the strengths from each physical component in 
Table~\ref{tab: components} and the off-diagonal ones describe their mixing (see App.~\ref{app: ncmec} for details).
The results are given in Fig.~\ref{fig: norm_contributions1}.

In general, we find significant and interesting differences between states with different flavour and isospin. 
In the $I(J^P)=0(1^+)$ tower, the $cc\bar{n}\bar{n}$ state (the $T_{cc}^+$) is clearly special as
it is almost purely determined by the attractive $DD^\ast$ contribution. This may be expected since its mass 
is extremely close to the $DD^\ast$ threshold and it is thus a prime candidate for a meson molecule. 
From our current formalism we cannot asses the spatial extent of this state and thus we can neither confirm 
nor discard the notion that the $D$ and $D^\ast$ components are far apart. Nevertheless it is striking that 
the underlying dynamics of dressed one-gluon exchange is capable of dynamically producing such a state. This 
is even more interesting as other flavour combinations show a completely different picture. The $bb\bar{n}\bar{n}$ 
bottom partner (the $T_{bb}^-$) is much more diverse.
Here we find a dominant attractive diquark $A_{bb}S$ component with about 32\% (dark green) contribution,  followed by 
its mixing with the attractive $BB^\ast$ component (19\%, light blue) and the $BB^\ast$ contribution itself
(18\%, dark blue). The other components are all below 10\%. 
These findings qualitatively agree with lattice studies \cite{Bicudo2021} (although it may not be entirely clear 
whether the mixings are directly comparable).
The situation changes again for the $T_{bc}$ with $bc\bar{n}\bar{n}$, which is strongly dominated 
by the attractive $\bar{B}^\ast D$ component (70\%, red), whose mixing with the attractive $A_{bc}S$ 
component (18\%, orange red) is also quite prominent. 
This  pattern is different from both $bb$ and $cc$ and can be attributed to the
different symmetries in the $bc$ case~\cite{Garcilazo2020,Deng2022},
which we  confirm  dynamically from the underlying QCD interaction.

It is also striking to see the effect of changed symmetries in the different isospin channels: Whereas in the $I(J^P)=0(1^+)$ tower we find significant differences between  $T_{cc}^+$ and $T_{bb}^-$, such differences are 
absent for $I(J^P)=1(1^+)$. Both $QQ \bar{n}\bar{n}$ states in this case are dominated by the mixing effects of 
the meson-meson component with the attractive diquark components. 

\section{Conclusions}\label{conclusions} 

In this work we studied the masses and binding energies of 
open-flavour four-quark states with two heavy quarks, $Q,Q' \in \{b,c\}$, 
using functional methods in QCD, employing Dyson-Schwinger equations and a four-body Faddeev-Yakubowski equation. 
{Assuming dominance of two-body correlations},
we demonstrated that the interactions between the four quarks  
can create very different internal structures depending on the  flavour content.
The $T^+_{cc}$ is completely dominated by $DD^\ast$ components due to its proximity to the respective threshold,
whereas the $T_{bb}^-$ is deeply bound and its internal structure is a complicated mixture of meson-meson and diquark components.
The  pattern with variations of flavour and isospin, discussed around Fig.~\ref{fig: norm_contributions1}, emerges from the same underlying 
interaction, namely dressed gluon exchange {in rainbow-ladder approximation}, which dynamically creates meson 
and diquark clusters without further assumptions. 
The quantitative results for the binding energies are highly non-trivial
and very encouraging results in our approach. 
Similar analyses for hidden-flavour and other open-flavour states are subject to future work {as well as further systematic
studies regarding the quality of our approximations}.

\acknowledgments{We thank Marc Wagner and the Frankfurt group for pointing out the importance of the repulsive colour channels 
and for extended and very fruitful discussions on the subject. This work was supported by the BMBF under project number 05P2021, 
the DFG under grant number FI 970/11-2, the graduate school HGS-HIRe and the GSI Helmholtzzentrum f\"ur Schwerionenforschung. 
This work contributes to the aims of the U.S. Department of Energy ExoHad Topical Collaboration, contract DE-SC0023598. 
We acknowledge computational resources provided by the HPC Core Facility and the HRZ of the Justus-Liebig-Universit\"at Gie\ss en.}

\clearpage

\appendix

\section{Construction of open-flavour amplitudes for unequal quarks}\label{app: new amplitudes}

Open-flavour four-quark states with quark content $QQ'\bar{q}\bar{q}'$ do not obey charge-conjugation symmetry, but rather the wave function of the state is subject to Pauli antisymmetry under the exchange of $Q\leftrightarrow Q'$ or $\bar{q}\leftrightarrow\bar{q}'$. The tensor bases for the quantum numbers investigated in this work were constructed accordingly and can be found in the Supplemental Material of \cite{Wallbott2020}. However, these works only considered open-flavour states with equal heavy-quark pairs $QQ'$. For an unequal heavy-quark pair $bc$, the Pauli antisymmetry does not hold when exchanging $b$ and $c$ quarks. Therefore, these bases need to be  modified  to investigate these states. 

Given the necessary Dirac-colour tensors to construct an axialvector four-quark state with $J^P=1^+$, i.e., Eq.~(A10) in the supplemental material of \cite{Wallbott2020}, one can arrange these tensors in four `categories', denoted by $\Psi_{(s/a)(s/a)}$. Each fulfils a certain Pauli symmetry ($s$) or antisymmetry ($a$) when exchanging $Q\leftrightarrow Q'$ or $\bar{q}\leftrightarrow\bar{q}'$ depending on the flavour wave function. These are: i) both Pauli symmetric $\left(\Psi_{ss}\right)$, ii) both Pauli antisymmetric $\left(\Psi_{aa}\right)$, iii) symmetric under $Q\leftrightarrow Q'$ and antisymmetric under $\bar{q}\leftrightarrow\bar{q}'$ $\left(\Psi_{sa}\right)$, and iv) antisymmetric under $Q\leftrightarrow Q'$ and symmetric under $\bar{q}\leftrightarrow\bar{q}'$ $\left(\Psi_{as}\right)$. To make a state with equal heavy-quark pair and $I(J^P)=0(1^+)$, one needs the tensors to be antisymmetric when exchanging the heavy-quark pair and symmetric when exchanging the light, i.e., $\Psi_{as}$. This then yields the components $DD^\ast$, $D^\ast D^\ast$ and $A_{cc}S$ for the $cc\bar{q}\bar{q}$ and $BB^\ast$, $B^\ast B^\ast$ and $A_{bb}S$ for the $bb\bar{q}\bar{q}$ state.
However, for an unequal heavy-quark pair there is no symmetry when exchanging $Q\leftrightarrow Q'$, whereas the symmetry in $\bar{q}\leftrightarrow\bar{q}'$ is still present. One can then form a linear combination of the tensors in $\Psi_{as}$ with the tensors in $\Psi_{ss}$, effectively destroying the (anti)symmetry in the first index of $\Psi$ while retaining the symmetry in the second index, i.e., $\bar{q}\leftrightarrow\bar{q}'$.
This disentangles the basis elements for the pseudoscalar-vector heavy-light meson-meson components, which then yields the following components for the $0(1^+)$ with quark content $bc\bar{q}\bar{q}'$: $\bar{B}D^\ast$, $\bar{B}^\ast D$, $\bar{B}^\ast D^\ast$ and $A_{bc}S$.
We neglect the attractive $\bar{B}^\ast D^\ast$ in the physical components in Table~\ref{tab: components} as it is the highest of the heavy-light meson-meson thresholds and was found to have negligible effect on the mass. However, for the isospin 
$I=1$ $bc\bar{q}\bar{q}$ states, the repulsive $\bar{B}^\ast D^\ast$ component was found to significantly contribute to the mass, pushing it into the correct region.

A further note has to be made regarding the colour basis elements. The respective combinations that make up an overall colour singlet are given in the Supplemental Material of \cite{Wallbott2020}.
{From $\mathbf{3}\otimes \mathbf{3} \otimes \mathbf{\bar{3}} \otimes \mathbf{\bar{3}} =
( \mathbf{\bar{3}} \oplus \mathbf{6}) \otimes (\mathbf{3}\oplus\mathbf{\bar{6}}) = \mathbf{1} \oplus \mathbf{1} \oplus ... $, the
colour part of the amplitude
consists of two independent colour singlet tensors.}
 These include the attractive meson-meson {colour wave functions}
\begin{align*}
C_{11} = \frac{1}{3}\,\delta_{AC}\delta_{BD},\qquad C'_{11} = \frac{1}{3}\,\delta_{AD}\delta_{BC}\, ,
\end{align*}
the attractive and repulsive diquark-antidiquark {colour wave functions}
\begin{align*}
C_{\bar{3}3} = -\frac{\sqrt{3}}{2}\left(C_{11}-C'_{11}\right), \quad C_{6\bar{6}} = \sqrt{\frac{3}{8}}\left(C_{11}+C'_{11}\right)\, ,
\end{align*}
and the repulsive meson-meson
\begin{align*}
C_{88} = \frac{C_{11}-3 C'_{11}}{2\sqrt{2}},\qquad C'_{88} = \frac{C'_{11}-3 C_{11}}{2\sqrt{2}} 
\end{align*}
components, where  $A,B,C,D=1,2,3$ denote the colour indices.
Under Pauli symmetry, these tensors transform as follows:
\begin{alignat}{6}
C_{11} &&\leftrightarrow C'_{11}\,, \qquad &C_{88} &&\leftrightarrow C'_{88} \,,\label{eq: symmMeson}\\[4pt]
C_{\bar{3}3} &&\leftrightarrow -C_{\bar{3}3}\,, \qquad & C_{6\bar{6}} &&\leftrightarrow C_{6\bar{6}}\,. \label{eq: symmdq}
\end{alignat}
While the meson-meson component stays the same\footnote{For the $1(1^+)$ $bc\bar{q}\bar{q}$ and $0(1^+)$ $bc\bar{s}\bar{s}$ states the repulsive $\bar{B}^\ast D^\ast$ component was found to be more significant than the repulsive $\bar{B}^\ast D$ component.} in Table~\ref{tab: components} for the attractive and repulsive channels due to Eq.~(\ref{eq: symmMeson}), the different sign in the symmetry transformations of Eq.~(\ref{eq: symmdq})  forces the repulsive diquark-antidiquark Dirac-colour tensors to fulfil the `mirrored' Pauli symmetry from the attractive diquark-antidiquark tensors, e.g., $\Psi_{as}$ for attractive $\to$ $\Psi_{sa}$ for the repulsive channel. Therefore, different diquark-antidiquark components appear in $f_2$ and $f_5$ in Table~\ref{tab: components}.

\begin{table}  \renewcommand{\arraystretch}{1.2}
	\setlength{\tabcolsep}{0.5em} 
	\centering
	\begin{tabular}{C{4mm}|C{5mm}C{5mm}C{8mm}|C{5mm}C{5mm}C{8mm}||C{5mm}C{5mm}}
		& \multicolumn{3}{c|}{$0^{-+}$} & \multicolumn{3}{c||}{$1^{--}$} & $0^+$ & $1^+$\\[4pt]
		& PDG & $m_{\rm{RL}}$ & $\Delta m^{\rm{rel.}}$ & PDG & $m_{\rm{RL}}$ & $\Delta m^{\rm{rel.}}$ & $m_{0^+}$ & $m_{1^+}$ \\[1mm]
		\hline\hline\rule{-0.9mm}{4mm}
		$n\bar{n}$ &  $\pi/\eta\, ^\dagger$ &  $\num{137}$ & $0.0\%$ & $\rho/\omega$ & $736$ & $5.2\%$ & $809$ & $1006$ \\
		$s\bar{n}$ &  $K$ & $501$ &$1.1\%$ & $K^\ast$ & $913$ & $0.1\%$ & $1072$ & $1259$  \\
		$s\bar{s}$ &  $-$ & $698$ &$-$ & $\phi$ & $1070$ & $5.0\%$ & $1266$ & $1412$ \\
		$c\bar{n}$ &  $D$ & $1860$ &$0.4\%$ & $D^\ast$ & $2011$ & $0.1\%$ & $2421$ & $2439$ \\
		$c\bar{s}$ &  $D_s$ & $1937$ &$1.6\%$ & $D_s^\ast$ & $2124$ & $0.5\%$ & $2523$ & $2543$  \\
		$c\bar{c}$ &  $\eta_c$ & $2803$ &$6.1\%$ & $J/\psi$ & $2992$ & $3.4\%$ & $3415$ & $3433$  \\
		$b\bar{n}$ &  $B$ & $5310$ &$0.6\%$ & $B^\ast$ & $5375$ & $0.9\%$ & $6396$ & $6403$  \\
		$b\bar{s}$ &  $B_s$ & $5425$ &$1.1\%$ & $B_s^\ast$ & $5487$ & $1.3\%$ &  $6473$ & $6492$ \\
		$b\bar{c}$ &  $B_c$ & $6232$ &$0.7\%$ & $-$ & $6302$ & $-$ & $7139$ & $7269$   \\
		$b\bar{b}$ &  $\eta_b$ & $9421$ & $0.2\%$ & $\Upsilon$ & $9500$ & $0.4\%$ & $9915$ & $10394$  \\
	\end{tabular}
	\caption{\label{tab: meson-masses}
		$Q\bar{q}$ mesons with quantum numbers $J^{PC} = \{0^{-+},1^{--}\}$ grouped according to their quark model classification.
		We show the experimental candidates \cite{Workman2022}, the masses $m_{\rm{RL}}$ obtained
		in our rainbow-ladder calculation, and the relative error of these to the masses given in the particle data group (PDG) (if the experimental state has been identified).
		In the last two columns we also show the obtained rainbow-ladder masses for the corresponding $Qq$ diquarks with quantum numbers $J^P = \{0^+,1^+\}$.
		All values are given in MeV. 
		$\dagger:$ The $\pi$ and the $\eta$ are mass degenerate in this work, since we neglect the strange component in the $\eta$ and the indirect
		effect of the topological mass via octet-singlet mixing. 
	}
\end{table}

\section{Two-body interaction}\label{app: two-body}

\begin{table}[h!]
	\setlength{\tabcolsep}{0.60em} 
	\centering
	\begin{tabular}{c|c|c|c|c}
		&  \multicolumn{2}{c|}{$0(1^{+})$} & \multicolumn{2}{c}{$1(1^{+})$} \\[4pt]
		\rule{0pt}{1.\normalbaselineskip}& $M$ & $E_B$& $M$ & $E_B$ \\
		\hline\hline\rule{-0.9mm}{4mm}
		\rule{0pt}{1.\normalbaselineskip}
		$cc\bar{n}\bar{n}$ & $3.89(5)$ & \unbound{$0.02(5)$} & $4.03(12)$ & \unbound{$0.16(12)$}\\[4pt]
		$cc\bar{s}\bar{s}$ & $4.23(10)$ & \unbound{$0.17(10)$} & $-$ & $-$\\[4pt]
		$bb\bar{n}\bar{n}$ & $10.57(4)$ & $-0.11(4)$ & $10.71(11)$ & \unbound{$0.03(11)$}\\[4pt]
		$bb\bar{s}\bar{s}$ & $10.83(10)$ & $-0.08(10)$ & $-$ & $-$\\[4pt]
		$bb\bar{c}\bar{c}$ & $12.20(9)$ & $-0.33(9)$ & $-$ & $-$
	\end{tabular} \\[5mm] 
	%
	\begin{tabular}{c|c|c|c|c}
		&   \multicolumn{2}{c|}{$0(1^{+})$} & \multicolumn{2}{c}{$1(1^{+})$} \\[4pt]
		\rule{0pt}{1.\normalbaselineskip}& $M$ & $E_B$& $M$ & $E_B$ \\
		\hline\hline\rule{-0.9mm}{4mm}
		\rule{0pt}{1.\normalbaselineskip}
		$cc\bar{n}\bar{n}$ & $4.43(3)$ & \unbound{$0.56(3)$} & $4.41(9)$ & \unbound{$0.54(9)$}\\[4pt]
		$cc\bar{s}\bar{s}$ & $4.61(8)$ & \unbound{$0.55(8)$} & $-$ & $-$\\[4pt]
		$bb\bar{n}\bar{n}$ & $10.92(15)$ & \unbound{$0.24(15)$} & $11.27(10)$ & \unbound{$0.58(10)$}\\[4pt]
		$bb\bar{s}\bar{s}$ & $11.40(8)$ & \unbound{$0.49(8)$} & $-$ & $-$\\[4pt]
		$bb\bar{c}\bar{c}$ & $12.77(7)$ & \unbound{$0.23(7)$} & $-$ & $-$
	\end{tabular}
	\caption{\label{tab: masses} Ground-state masses (upper panel) and first radial excitations (lower panel) for the open-charm ($cc\bar{q}\bar{q}$) and open-bottom ($bb\bar{q}\bar{q}$) states in GeV.
		For completeness we also display the binding energies $E_B$ with respect to the  lightest (calculated) heavy-light meson-meson threshold in each channel; the ``binding energies''
		for resonant particles above the threshold are shown in grey. The error given in the brackets is the combination of the extrapolation error
		and the error of the fit to the quark mass evolution curve.}
%
	\setlength{\tabcolsep}{0.60em} 
	\centering
	\begin{tabular}{c|c|c|c|c}
		&  \multicolumn{2}{c|}{$0(1^{+})$} & \multicolumn{2}{c}{$1(1^{+})$} \\[4pt]
		\rule{0pt}{1.\normalbaselineskip}& $M$  & $E_B$ & $M$ & $E_B$ \\
		\hline\hline\rule{-0.9mm}{4mm}
		\rule{0pt}{1.\normalbaselineskip}
		$bc\bar{n}\bar{n}$ & $7.24(3)$ & \unbound{$0.00(3)$} & $7.45(5)$ & \unbound{$0.22(5)$}\\[4pt]
$bc\bar{s}\bar{s}$ & $7.61(5)$ & $\unbound{0.19(5)}$ & $-$ & $-$\\[4pt]
$bc\bar{c}\bar{c}$ & $9.02(1)$ & $-0.09(1)$ & $-$ & $-$\\[4pt]
$bc\bar{b}\bar{b}$ & $14.86(1)$ & $-0.88(1)$ & $-$ & $-$\\[4pt]
	\end{tabular}\\[5mm]
	
	\begin{tabular}{c|c|c|c|c}
		&  \multicolumn{2}{c|}{$0(1^{+})$} & \multicolumn{2}{c}{$1(1^{+})$} \\[4pt]
		\rule{0pt}{1.\normalbaselineskip}& $M$  & $E_B$ & $M$ & $E_B$ \\
		\hline\hline\rule{-0.9mm}{4mm}
		\rule{0pt}{1.\normalbaselineskip}
		$bc\bar{n}\bar{n}$ & $7.38(3)$ & \unbound{$0.15(3)$} & $7.52(8)$ & \unbound{$0.29(8)$}\\[4pt]
$bc\bar{s}\bar{s}$ & $7.68(8)$ & \unbound{$0.26(8)$} & $-$ & $-$\\[4pt]
$bc\bar{c}\bar{c}$ & $9.09(6)$ & $-0.01(6)$ & $-$ & $-$\\[4pt]
$bc\bar{b}\bar{b}$ & $14.94(0)$ & $-0.79(0)$ & $-$ & $-$\\[4pt]
	\end{tabular}
	\caption{\label{tab: masses_unequal} Ground-state and excited state masses in GeV for four-quark states $QQ'\bar{q}\bar{q}'$ with unequal heavy $QQ'$ pair. We also display the binding energies with respect to the lowest relevant meson-meson thresholds, i.e., $B^\ast D$ for the $1^+$ state. Masses of resonant particles are above the thresholds and their ``binding energies'' are shown in grey.}
\end{table}

We employ a rainbow-ladder truncation combined with the effective Maris-Tandy (MT) interaction \cite{Maris1997,Maris1999} for the two-body interaction kernels, which has proven to be a very successful approximation in the functional framework. For recent reviews on the application to the meson, baryon and four-quark phenomenology in the past, see \cite{Eichmann2016,Eichmann2020}. This interaction has also been used to compute the quark Dyson-Schwinger equation (DSE) and the meson and diquark mass spectrum obtained from the quark-(anti)quark BSE (see Table~\ref{tab: meson-masses}), both of which are needed as input for the four-quark BSE. The interaction in its explicit form can be found in Eq.~(3.96) of \cite{Eichmann2016}; in this work we use the scale parameter $\Lambda = 0.72$ tuned to reproduce the pion decay constant and fix the shape parameter to $\eta^{\rm{MT}} = 1.8$. The MT interaction describes the phenomenology of light mesons in the pseudoscalar and vector channel reasonably well, whereas the results in the scalar and axialvector channels are not satisfactory. The qualitative reliability of the interaction can be judged from the two-body meson masses {displayed in Table I in Ref.\cite{Hoffer2024}.  
Herein, for the convenience of the reader, we display in Table~\ref{tab: meson-masses} again the masses of the 
pseudoscalar and vector mesons that enter our calculation.}

\begin{figure}
	\centering
	\includegraphics[width=1.0\columnwidth]{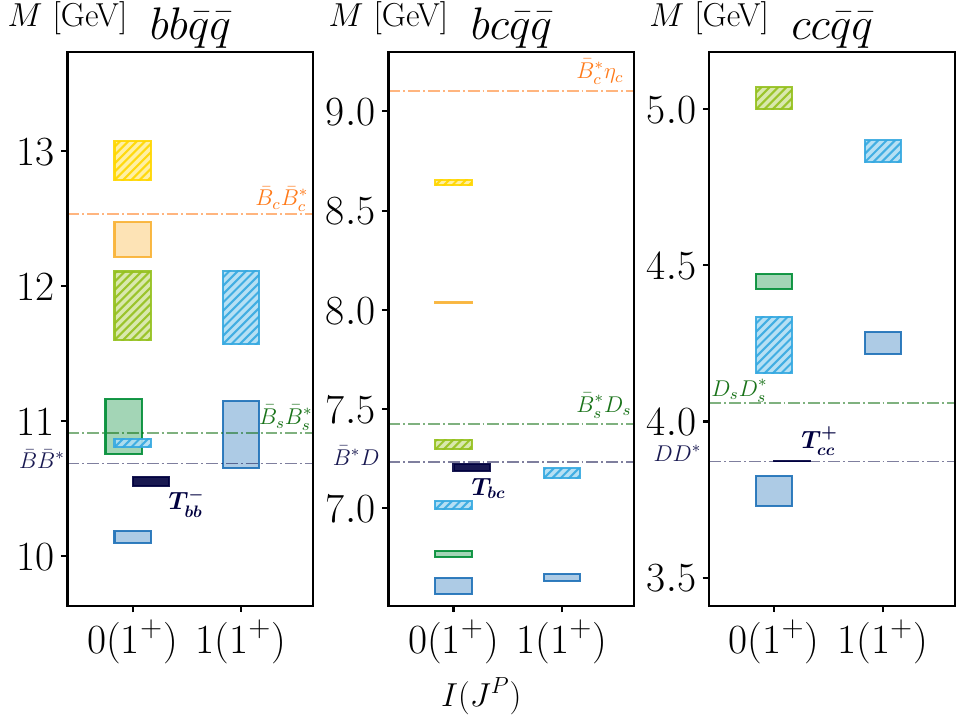}
	\caption{\label{fig:spectrum_att_only} Same plot as in Fig.~\ref{fig: equal_QQ}, but here we 
		show the spectra obtained by only including the attractive components given in 
		Table~\ref{tab: components}.}
\end{figure}

At the quark level, and consequently also at the meson level, we work in the isospin symmetric limit, i.e., 
$m_{\pi^\pm} = m_{\pi^0}$, $m_{D^\pm} = m_{D^0}$, $m_{B^\pm}=m_{B^0}$. 
The $u/d$ current-quark mass is fixed by $m_\pi$. The strange, charm and bottom quark masses 
are fixed such that the following criteria are satisfied: the sums $m_D+m_{D^\ast}$, $m_{D_s}+m_{D_s^\ast}$ and $m_B+m_{B^\ast}$ 
have to match the sum of the respective experimental values \cite{Workman2022} within $1 \%$ relative error. Note, however, that the isospin limit
is broken at the level of the four-point functions by symmetries and the potential appearance of 
attractive and repulsive channels displayed in Table \ref{tab: components}.

\section{Masses and binding energies of open flavour four-quark states}\label{app: masses}

In Tables \ref{tab: masses} and \ref{tab: masses_unequal} we give explicit numbers for the masses and 
binding energies of the states discussed in the main body of this work in Sec.~\ref{results}.

\begin{table}[t!]
	\setlength{\tabcolsep}{0.60em} 
	\centering
	\begin{tabular}{c|c|c|c|c}
		&  \multicolumn{2}{c|}{$0(1^{+})$} & \multicolumn{2}{c}{$1(1^{+})$} \\[4pt]
		\rule{0pt}{1.\normalbaselineskip}& $M$ & $E_B$& $M$ & $E_B$ \\
		\hline\hline\rule{-0.9mm}{4mm}
		\rule{0pt}{1.\normalbaselineskip}
		$cc\bar{n}\bar{n}$ & $3.78(5)$ & $-0.09(5)$ & $4.25(3)$ & \unbound{$0.38(3)$}\\[4pt]
		$cc\bar{s}\bar{s}$ & $4.45(2)$ & \unbound{$0.39(2)$} & $-$ & $-$\\[4pt]
		$bb\bar{n}\bar{n}$ & $10.14(4)$ & $-0.54(4)$ & $10.90(25)$ & \unbound{$0.21(25)$}\\[4pt]
		$bb\bar{s}\bar{s}$ & $10.96(20)$ & \unbound{$0.05(20)$} & $-$ & $-$\\[4pt]
		$bb\bar{c}\bar{c}$ & $12.35(13)$ & $-0.19(13)$ & $-$ & $-$
	\end{tabular} \\[5mm] 
	%
	\begin{tabular}{c|c|c|c|c}
		&   \multicolumn{2}{c|}{$0(1^{+})$} & \multicolumn{2}{c}{$1(1^{+})$} \\[4pt]
		\rule{0pt}{1.\normalbaselineskip}& $M$ & $E_B$& $M$ & $E_B$ \\
		\hline\hline\rule{-0.9mm}{4mm}
		\rule{0pt}{1.\normalbaselineskip}
		$cc\bar{n}\bar{n}$ & $4.25(9)$ & \unbound{$0.37(9)$} & $4.87(4)$ & \unbound{$0.99(4)$}\\[4pt]
		$cc\bar{s}\bar{s}$ & $5.04(3)$ & \unbound{$0.97(3)$} & $-$ & $-$\\[4pt]
		$bb\bar{n}\bar{n}$ & $10.84(3)$ & \unbound{$0.15(3)$} & $11.84(27)$ & \unbound{$\unbound{1.16(27)}$}\\[4pt]
		$bb\bar{s}\bar{s}$ & $11.85(25)$ & $\unbound{0.94(25)}$ & $-$ & $-$\\[4pt]
		$bb\bar{c}\bar{c}$ & $12.93(15)$ & $\unbound{0.40(15)}$ & $-$ & $-$
	\end{tabular}
	\caption{\label{tab: masses_att} Masses and binding energies with respect to the lowest meson-meson threshold for the ground-state (upper panel) and first radial excitations (lower panel) for the open-charm ($cc\bar{q}\bar{q}$) and open-bottom ($bb\bar{q}\bar{q}$) states obtained by using the attractive colour components only. The corresponding spectrum is shown in Fig.~\ref{fig:spectrum_att_only}. Masses are given in GeV.
		}
%
	\setlength{\tabcolsep}{0.60em} 
	\centering
	\begin{tabular}{c|c|c|c|c}
		&  \multicolumn{2}{c|}{$0(1^{+})$} & \multicolumn{2}{c}{$1(1^{+})$} \\[4pt]
		\rule{0pt}{1.\normalbaselineskip}& $M$  & $E_B$ & $M$ & $E_B$ \\
		\hline\hline\rule{-0.9mm}{4mm}
		\rule{0pt}{1.\normalbaselineskip}
		$bc\bar{n}\bar{n}$ & $6.61(4)$ & $-0.63(4)$ & $6.65(2)$ & $-0.58(2)$\\[4pt]
		$bc\bar{s}\bar{s}$ & $6.77(2)$ & $-0.65(2)$ & $-$ & $-$\\[4pt]
		$bc\bar{c}\bar{c}$ & $8.04(0)$ & $-1.07(0)$ & $-$ & $-$\\[4pt]
		$bc\bar{b}\bar{b}$ & $14.01(1)$ & $-1.72(1)$ & $-$ & $-$\\[4pt]
	\end{tabular}\\[5mm]
	
	\begin{tabular}{c|c|c|c|c}
		&  \multicolumn{2}{c|}{$0(1^{+})$} & \multicolumn{2}{c}{$1(1^{+})$} \\[4pt]
		\rule{0pt}{1.\normalbaselineskip}& $M$  & $E_B$ & $M$ & $E_B$ \\
		\hline\hline\rule{-0.9mm}{4mm}
		\rule{0pt}{1.\normalbaselineskip}
		$bc\bar{n}\bar{n}$ & $7.02(2)$ & $-0.22(2)$ & $7.18(2)$ & $-0.06(2)$\\[4pt]
		$bc\bar{s}\bar{s}$ & $7.32(2)$ & $-0.10(2)$ & $-$ & $-$\\[4pt]
		$bc\bar{c}\bar{c}$ & $8.64(1)$ & $-0.46(1)$ & $-$ & $-$\\[4pt]
		$bc\bar{b}\bar{b}$ & $14.53(1)$ & $-1.20(1)$ & $-$ & $-$\\[4pt]
	\end{tabular}
	\caption{\label{tab: masses_unequal_att} Ground-state and excited state masses in GeV for four-quark states $QQ'\bar{q}\bar{q}'$ with unequal heavy $QQ'$ pair when only using the attractive colour channels. We also display the binding energies with respect to the lowest relevant meson-meson thresholds, i.e., $B^\ast D$ for the $1^+$ state. Masses of resonant particles are above the thresholds and their ``binding energies'' are shown in grey.}
\end{table}

\begin{figure*}[t]
	\centering
	\includegraphics[width=1.0\textwidth]{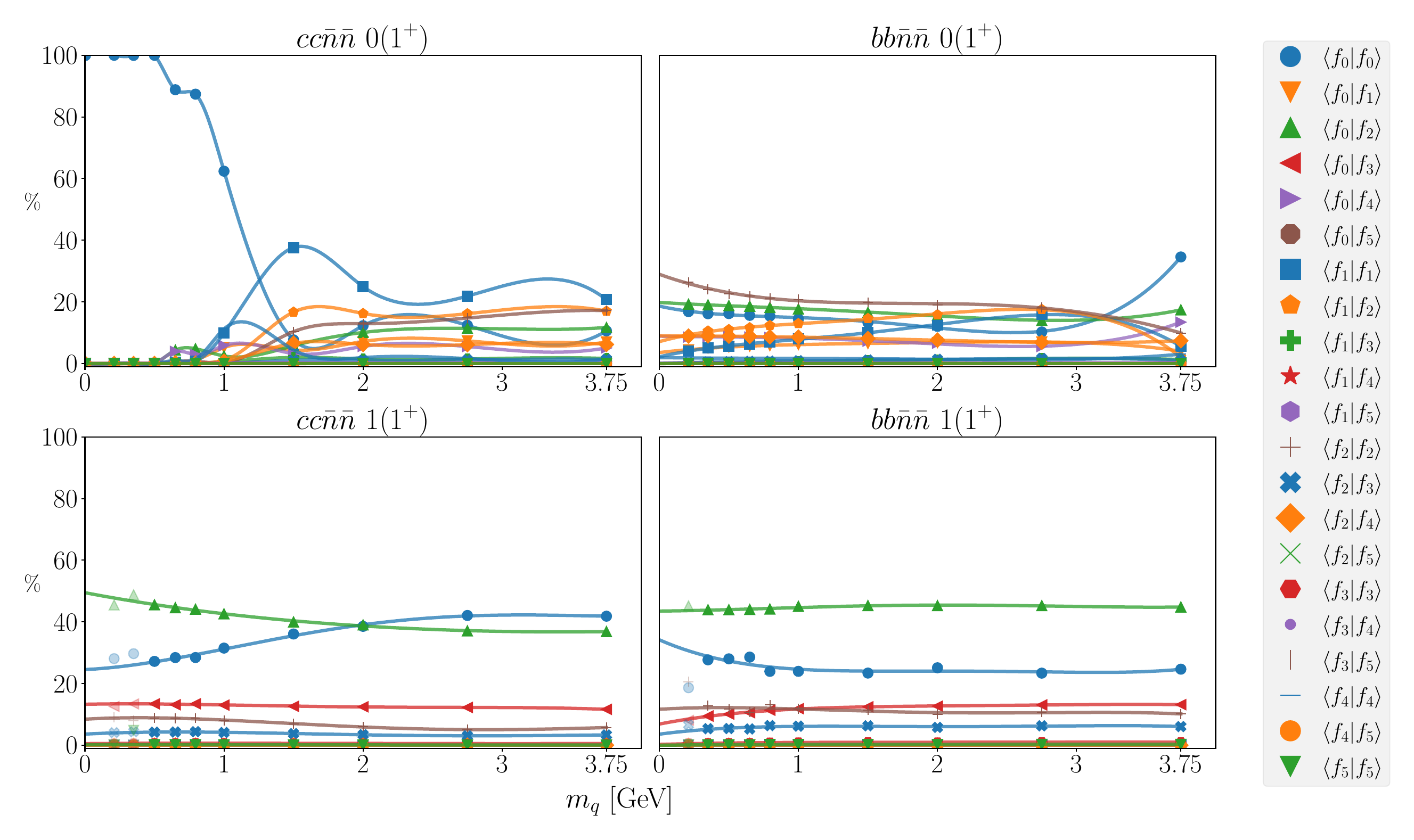}
	\caption{\label{fig: ncmec_eq_QQ}Quark mass evolution of the norm contributions for the equal heavy-quark pair ground states in the $0(1^+)$ (\textit{top row}) and the $1(1^+)$ (\textit{bottom row}) channel. The left panels show the charm ($cc\bar{q}\bar{q}$) and the right panels the bottom ($bb\bar{q}\bar{q}$) states. 
		The colours of the fits correspond to the fitted data points. Data points depicted as opaque were not taken into account in the fits.  }
	\end{figure*}
	\begin{figure*}[t]
	\centering
	\includegraphics[width=1.0\textwidth]{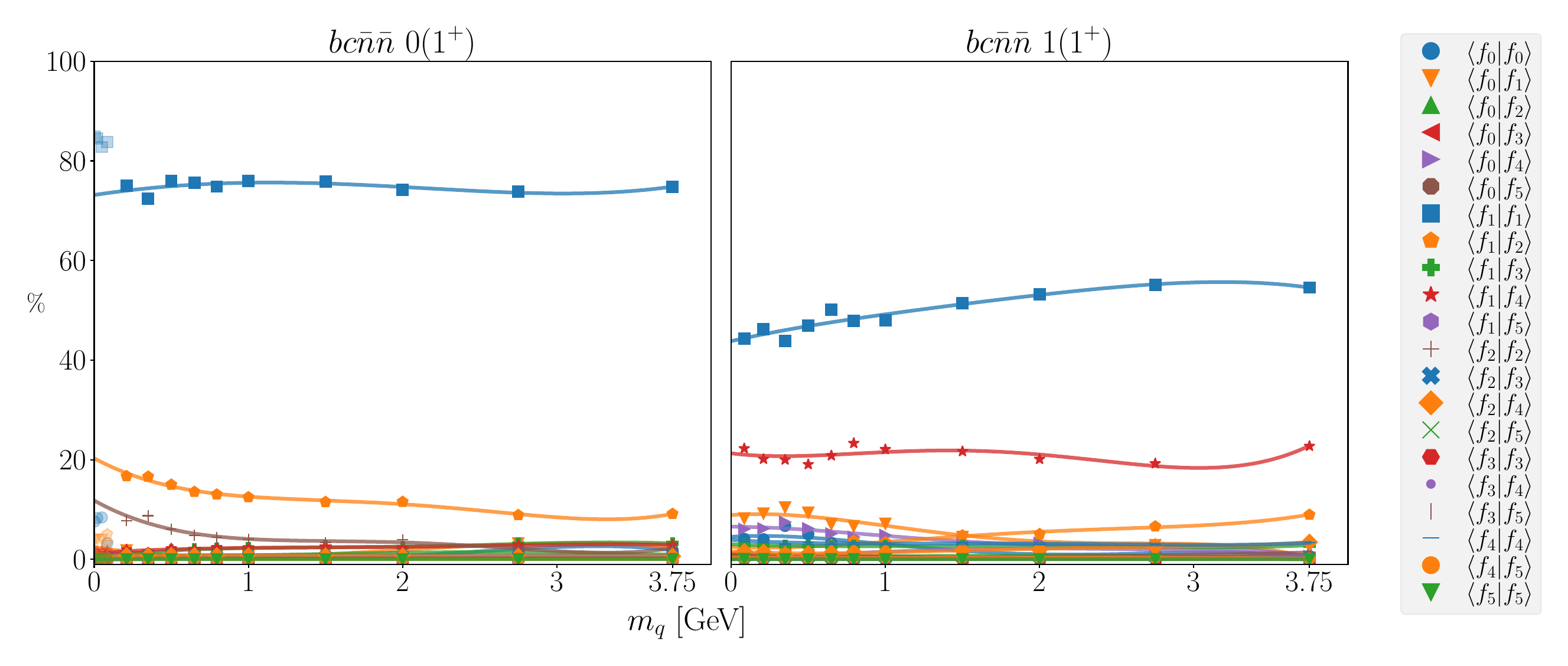}
	\caption{\label{fig: ncmec_uneq_QQ}Norm contribution quark mass evolution curves for the unequal heavy-quark pair ground states in the $0(1^+)$ and $1(1^+)$ channels. The colours of the fits correspond to the fitted data points. Data points depicted as opaque were not taken into account in the fits.}
\end{figure*}

	\section{Open-flavour spectrum with only attractive colour forces}\label{app: att_only}

{
	In Figure~\ref{fig:spectrum_att_only} we show the spectrum for the $J^P=1^+$ states where we only took the attractive colour components in Tab.~\ref{tab: components} into account. 
	The corresponding numerical values are given in Tab.~\ref{tab: masses_att} and Tab.~\ref{tab: masses_unequal_att}. We can now compare this spectrum to the one in Fig.~\ref{fig: equal_QQ}, where the attractive and repulsive colour components were used. 

Starting with the open-bottom spectrum, we find that the mass of the $bb\bar{n}\bar{n}$ ground state in the $0(1^+)$ channel moves up by about $400$ MeV when we include the repulsive colour 
channels. The mass of the corresponding excited state in this channel increases by 
about 800 MeV with the caveat that the errorbar here is quite large. By including the repulsive colour forces, the masses of the ground and excited states with $bb\bar{q}\bar{q}$ and $\bar{q}\neq \bar{n}$ can be determined more accurately compared 
to the attractive colour channels only case, leading to a reduction in error (cf. 
upper panels of Tab.~\ref{tab: masses} and Tab.~\ref{tab: masses_att}). Same goes for the 
$bb\bar{n}\bar{n}$ ground and excited state with $1(1^+)$. 

For the $bc\bar{q}\bar{q}$ states one can see that without the inclusion of the repulsive colour channels, the whole spectrum is found to be very deeply bound. Thus, the consideration of attractive and repulsive colour forces is of vital importance here, as the whole spectrum gets shifted upwards by $300$ to $900$ MeV into the correct mass region.

Regarding the open-charm spectrum, the $cc\bar{n}\bar{n}$ ground state in the $0(1^+)$ channel increases in mass by about $120$ MeV when including the repulsive colour channels. This moves the state right up to the $DD^\ast$ threshold, where it is 
expected from experiment \cite{Workman2022}.
}

As explained in the main part of this work, the results displayed in Fig.~\ref{fig:spectrum_att_only} have been obtained with improved
extrapolation methods as compared to the previous work by some of us, Ref.~\cite{Wallbott2020}. Thus they supersede those of 
Ref.~\cite{Wallbott2020} as concerns the open-flavour states.

\section{Norm contribution quark mass evolution curves}\label{app: ncmec}

To calculate the norm contributions in Fig.~\ref{fig: norm_contributions1}, the four-quark states must be on-shell. 
However, because the on-shell BSE is not always directly calculable due to internal two-body poles,
we vary the mass of the `light' $\bar{q}\bar{q}$ pair from $\bar{n}\bar{n}$ to $\bar{b}\bar{b}$ to determine the quark mass evolution of the norm contributions.
Their extrapolation to the physical $u/d$ quark mass then yields the result in Fig.~\ref{fig: norm_contributions1}.

Fig.~\ref{fig: ncmec_eq_QQ} shows the quark mass evolution of the norm contributions in the $0(1^+)$ (top row) and $1(1^+)$ (bottom row) channels
 for the $cc\bar{q}\bar{q}$ (left column) and  $bb\bar{q}\bar{q}$  states  (right column). 
 The analogous case for $bc\bar{q}\bar{q}$ with $0(1^+)$ is shown in Fig.~\ref{fig: ncmec_uneq_QQ}.
The different data points represent the correlations between the internal clusters given in Table~\ref{tab: components}, and the 
corresponding fits are coloured accordingly. To obtain each data point, we calculate the norm contribution 
at each $P^2$ in the eigenvalue curve and read off the result where the condition $\lambda(P^2)=1$ is satisfied.
For most states, the curves do not show much changes over the whole mass range from up/down to the bottom quark region. This is exceptionally
different for the $T_{cc}^+$, which shows dominance of the $DD^\ast$ for 'light' quark pairs only below about the charm quark mass $m_q<m_c$.

\clearpage

\bibliography{open_flavour}

\end{document}